\documentstyle[aps,preprint,epsf]{revtex}

\def\prl#1#2#3{Phys. Rev. Lett., {\bf #1}, #2 (#3)}
\def\pla#1#2#3{Phys. Letts. A {\bf #1}, #2 (#3)}
\def\pra#1#2#3{Phys. Rev. A, {\bf #1}, #2 (#3)}
\def\pre#1#2#3{Phys. Rev. E, {\bf #1}, #2 (#3)}
\def\phd#1#2#3{Physica D {\bf #1}, #2 (#3)}

\def\beq{\begin{equation}}
\def\eqn{\end{equation}\noindent}
\begin{document}
\title{Synchronization of Strange Nonchaotic Attractors}
\author{Ramakrishna Ramaswamy}
\address{School of Physical Sciences, Jawaharlal Nehru University, New
Delhi 110067, India}
\date{\today}
\maketitle
\begin{abstract}
Strange nonchaotic attractors (SNAs), which are realized in many
quasiperiodically driven nonlinear systems are strange (geometrically
fractal) but nonchaotic (the largest nontrivial Lyapunov exponent is
negative).  Two such identical independent systems can be synchronized
by in--phase driving: because of the negative Lyapunov exponent, the
systems converge to a common dynamics which, because of the strangeness
of the underlying attractor, is aperiodic. This feature, which is
robust to external noise, can be used for applications such as secure
communication. A possible implementation is discussed, and its
performance is evaluated. The use of SNAs rather than chaotic attractors
can offer some advantages in experiments involving
synchronization with aperiodic dynamics.
\end{abstract}
\vskip1cm

Pecora and Carroll \cite{PC} showed that identical (or nearly
identical) nonlinear systems can be made to synchronize if coupled by a
common drive signal. If one considers the overall system as separated
into drive and response subsystems, then a necessary and sufficient
condition for synchronization to occur is that the Lyapunov exponents
corresponding to the response subsytem are all negative.  This property
is robust, and is easy to realize in the laboratory
\cite{PC,C,GENERAL}, even when the dynamics of the drive is chaotic and
unstable. An application of chaotic synchronization that has been
extensively explored is the possibility of secure communications: a
number of different schemes based on a variety of coding principles
have been proposed \cite{CO,COMMUNICATIONS,ML,YYY}. 

The property of synchronization of nonlinear systems is extremely
general. One situation where this is most easily achieved is between
quasiperiodically driven systems in the regime wherein the dynamics
lies on strange nonchaotic attractors (SNAs) \cite{SNA,OTT}. The
purpose of this report is to suggest that such systems possess
advantages that make them ideal for applications in communications
which use aperiodic signals.

SNAs, which are found in quasiperiodically driven systems, are
geometrically strange, namely they are fractal, but the largest
nontrivial Lyapunov exponent is negative, and hence the dynamics is not
chaotic. They can be created through a variety of mechanisms
\cite{MECHANISMS}, and exist over a range of parameter values ({\it
i.e.} they are not exceptional or nongeneric).  SNAs have been observed
in several experimental systems \cite{DITTO,EXPERIMENTS}, and have been
verified through the use of power spectral methods and attractor
dimension estimates. Although the largest nontrivial Lyapunov exponent
is negative, the dynamics is {\it aperiodic} since the underlying
attractor is strange: this makes it difficult to deduce the Lyapunov
exponents, or indeed the {\it nonchaoticity}, by attractor
reconstruction using standard methods.

Synchronization of two such systems is trivial because of the negative
Lyapunov exponents. Regardless of where the systems are started, they
eventually converge to the same dynamics so long as the {\em phase} of
the quasiperiodic driving is matched. There is no requirement of
coupling the systems (other than the coupling implicit in the matched
phase; see below).  

As an example of this behaviour, consider the following system first
introduced by Zhou, Moss and Bulsara \cite{ZMB}, which describes a
driven-damped SQUID,
\beq
\ddot x + k \dot x = - (x+ \beta \sin 2 \pi x) + q_1 \sin \omega_1 t +
q_2 \sin \omega_2 t
\label{squid}
\eqn
where the ratio of frequencies is taken to be irrational, $\omega_1 /
\omega_2 = (\sqrt 5 +1 )/2$.
This system (and related variants) has been extensively studied in both
numerical as well as analog simulations, and is thus a typical example
of a system that can be experimentally realized. An identical copy of
this system  with phase--difference $\phi$ has the equation of motion
\beq
\ddot y + k \dot y = - (y+ \beta \sin 2 \pi y) + q_1 \sin \omega_1
(t+\phi) + q_2 \sin \omega_2 (t+\phi).
\label{resp}
\eqn
The two systems can be synchronized regardless of the initial values of
$x, \dot x, y, \dot y$, so long as there is no phase--lag, $\phi = 0$
and the parameters (here $k, \beta$ and $q_1, q_2$) are such that the
dynamics is on a SNA.  Explicitly, it is observed that $\vert x(t) -
y(t) \vert \to 0$ rapidly, and results for a typical orbit are shown in
Fig.~1a.

Rewriting the above system in autonomous form
\begin{eqnarray}
\dot x_1 &=& x_2 \nonumber\\
\dot x_2 &=& -k x_2 - (x_1 + \beta \sin 2 \pi x_1) + q_1 \sin \omega_1
x_3 + q_2 \sin \omega_2 x_3\nonumber\\
\dot x_3 &=& 1\nonumber\\
\dot y_1 &=& y_2\nonumber\\
\dot y_2 &=& -k y_2 - (y_1 + \beta \sin 2 \pi y_1) + q_1 \sin \omega_1
y_3 + q_2 \sin \omega_2 y_3 \label{note}\\
\dot y_3 &=& 1\nonumber
\end{eqnarray}\noindent
makes it evident that phase matching corresponds to replacing $y_3$ in
Eq.~(\ref{note}) by $x_3$ and thereby coupling the two systems.  This
then conforms to the general framework of synchronization in the manner
of Pecora and Carroll \cite{PC} with the `$x$' system the drive and the
`$y$' effectively the response.

In other parameter ranges, the system in Eq.~(\ref{squid}) can be
chaotic. In such a case, both the drive and the response have positive
Lyapunov exponents, and synchronization cannot occur---see (Fig.~1b).
When there is a phase mismatch, namely if $\phi \ne 0$ in
Eq.~(\ref{resp}), again synchronization does not occur (Fig.~1c), even
when the parameters correspond to SNA dynamics.

Secure communications using aperiodic dynamics has been implemented in
several ways \cite{CO,COMMUNICATIONS,ML,YYY}, and the technique of
synchronization with SNAs rather than chaotic attractors can be
employed in several of them. It should be mentioned, however, that some
of the simpler schemes have been shown susceptible to unmasking
\cite{PCP} by inference of the underlying attractor.

The most direct method of secure communication through the use of chaotic
synchronization uses a chaotic signal to mask information
\cite{CO}. The alternate strategy suggested here is to use the signal
from a strange nonchaotic system in an analogus manner, by transmitting
the low-amplitude information-bearing signal, $m(t)$ which is added to
(and masked by) the output from the first system, $x(t)$, namely
$x^{\prime}(t) = x(t)+m(t)$.  It is also necessary to simultaneously
transmit a means of phase-locking, say a train of $\delta-$function
pulses.  Recovery of $m(t)$ can be effected by allowing the systems to
synchronize and subtracting the output of the second system {\i.e.}
$x^{\prime}(t) - y(t)$.

Since the two systems evolve independently, the effect of additive noise
is minimal: noise added to $x^{\prime}(t)$ will be unchanged upon
subtraction. On the other hand, the mismatch between the transmitter
system and the response is necessary to consider in some detail.
One way to explore the effect of such mismatch is by introducing
fluctuations in the parameters of the response,
\beq
\mu = \mu_0 ( 1 + \sigma \xi (t))
\label{noiseq}
\eqn
where $\sigma$ is the noise amplitude and $\xi(t)$ is a
$\delta$--correlated random variable with zero mean, and $\mu_0$ is the
value of the parameter in the transmitter system. For the response
system in Eq.~(\ref{resp}), we consider $\mu \equiv q_2, \beta$ and
$\omega_2$. Results are shown in Fig.~2 for the case of noise amplitude
$\sigma$ = 10$^{-2}$ for the three parameters indicated above.  The
plot of $x$ {\it vs.} $y$ shows that the degree of synchronization in
the presence of noise is fairly good, except for the case of $\mu
\equiv \omega_2$, namely when the quasiperiodic driving frequency is
subject to fluctuations. Indeed, variation of the parameters $q_2$ and
$\beta$ by up to 10\% does not significantly alter the synchronization
except for short bursts in time. The drive frequency is much more
sensitive to fluctuations, and only by reducing the noise amplitude to
10$^{-4}$ is it possible to greatly improve the synchronization in this
case (Fig.~2d).

The viability of the above scheme is demonstrated using the SNA of
Eqs.~(\ref{squid}--\ref{resp}), and the results are shown in Fig.~3,
wherein the signal to be communicated is a sinusoidal form. In
the absence of noise, the recovery of the signal is exact (and is
therefore not shown); with noise added following Eq.~(\ref{noiseq}) in
the driving frequency or in the other parameters, the recovery of the
signal is of good quality. Indeed, even when the parameters of the two
systems do not match, signal recovery can be effected: in Fig.~3a the
parameters $q_2$ of the $x$ and $y$ systems differ by 10\%.  The
occasional errors due to loss of synchronization that are apparent in
Fig.~3 do not persist over long times.

Note that it is important for this means of application that
interception of $x^{\prime}(t)$ can have no potential value in the
absence of knowledge of the underlying dynamical system, since the
dynamics is intrinsically aperiodic. One can use standard methods to
reconstruct the dynamics \cite{WOLF}, but the extraction of reliable
values for (small) negative Lyapunov exponents from experimental
time--series data for SNAs has proven to be difficult
\cite{DITTO,EXPERIMENTS}.  Thus it may be more problematic to reliably
reconstruct the underlying attractor, in contrast to the example of the
Lorenz system which was considered by Perez and Cerdeira \cite{PCP}.

Related schemes that use chaotic attractors, as for example the
modulation/detection procedure described by Cuomo and Oppenheim
\cite{CO} can be similarly adapted to the case of SNAs. A somewhat
different implementation of secure communication using SNAs which
transmits digital information by switching parameter values has also
been proposed recently \cite{ZC}.

The synchronizing property arises directly from the use of a common
in--phase driving: the negative Lyapunov exponents alone do not
guarantee that the $x$ and $y$ signals will coincide. (In the extreme
case when both systems are integrable, in the absence of a common
driving term, there will be no synchronization.) Other applications
that use the synchronization of chaotic systems \cite{NATURE} can also
be effected using strange nonchaotic systems. In general, as a
consequence of the negative Lyapunov exponents, the stability and
robustness using SNAs is greater than that with comparable chaotic
attractors. This may make quasiperiodically driven systems particularly
suitable for applications that involve the synchronization of large
numbers of nonlinear dynamical systems.

\newpage

\begin{figure}[]
\epsfxsize=6cm
\epsfbox[20 350 200 650]{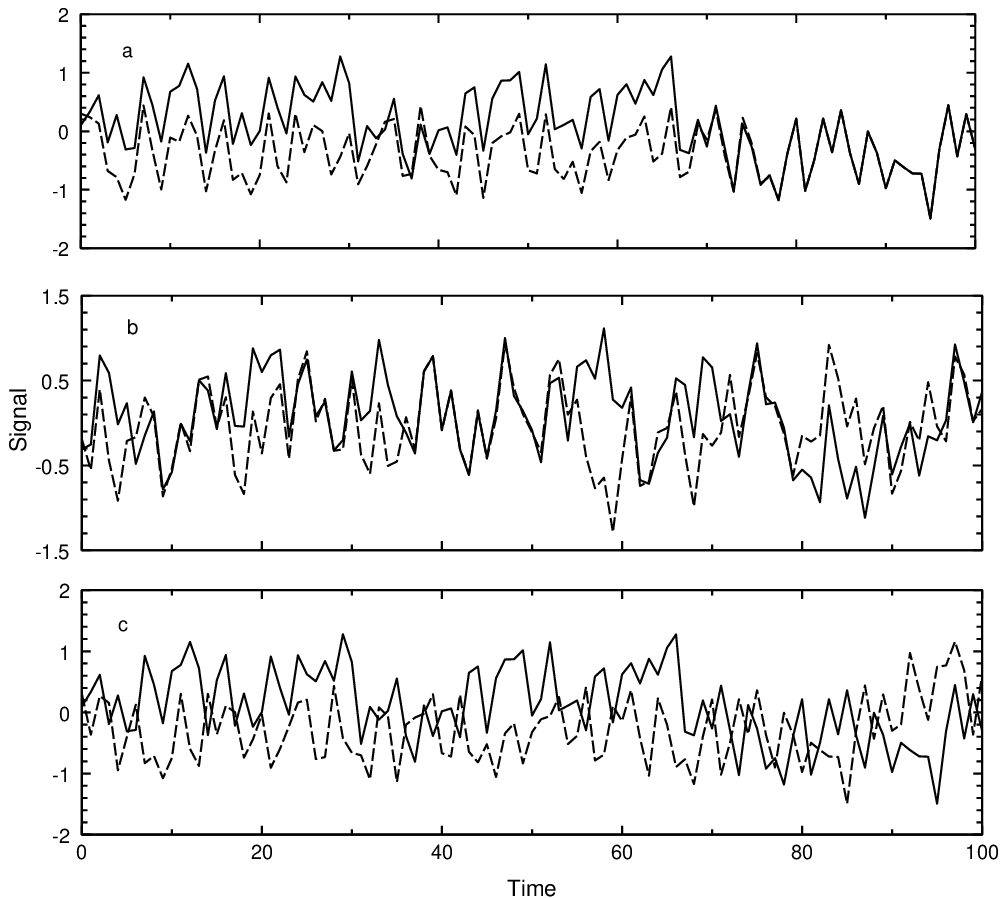}
\end{figure}  
\vskip5cm
Fig.~1 Time series of the signals from the two systems, $x_1(t)$ (solid line)
and $y_1(t)$ (dashed line).  The parameters are set at $k = \beta = 2,
q_1 = 2.768, \omega_1 = 2.25$. a) When $q_2 = 0.88$
and $\phi = 0$, the dynamics is on a SNA, and the two systems
synchronize.  b) When $q_2 = 0.38$ and $\phi = 0$, the dynamics is on a
chaotic attractor. The Lyapunov exponents are all positive, and
synchronization is not possible.  c) When $q_2 = 0.88$ and $\phi \ne
0$, the dynamics is on a SNA, but synchronization does not occur.

\newpage
\begin{figure}[]
\epsfxsize=6cm
\epsfbox[20 350 200 650]{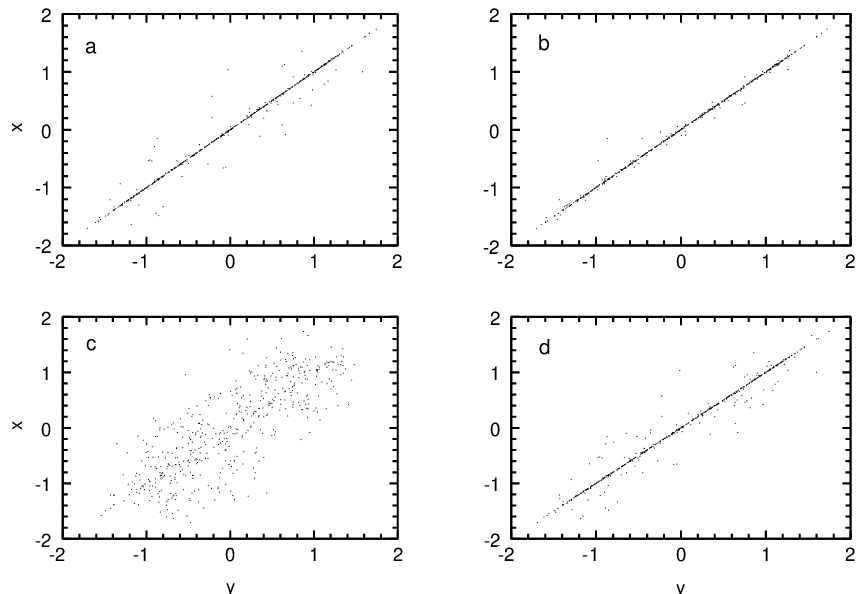}
\end{figure}  
\vskip6cm
Fig.~2 Robustness of synchronization with respect to noise which alters
the parameters of the response system, as described in the text. The
values of the parameters are $k = \beta = 2, q_1 = 2.768, q_2 = 0.88,
\omega_1 = 2.25$, and the noise strength is $\sigma
= 10^{-2}$, for a) $\mu \equiv \beta$ (See Eq.~\ref{noiseq}), b) $\mu
\equiv q_2$, and c) $\mu \equiv \omega_2$. Reduction of the noise
strength improves the synchronization in the last case, d) where
$\sigma = 10^{-4}$ and $\mu \equiv \omega_2$.

\newpage
\begin{figure}
\epsfxsize=4cm
\epsfbox[50 500 200 800]{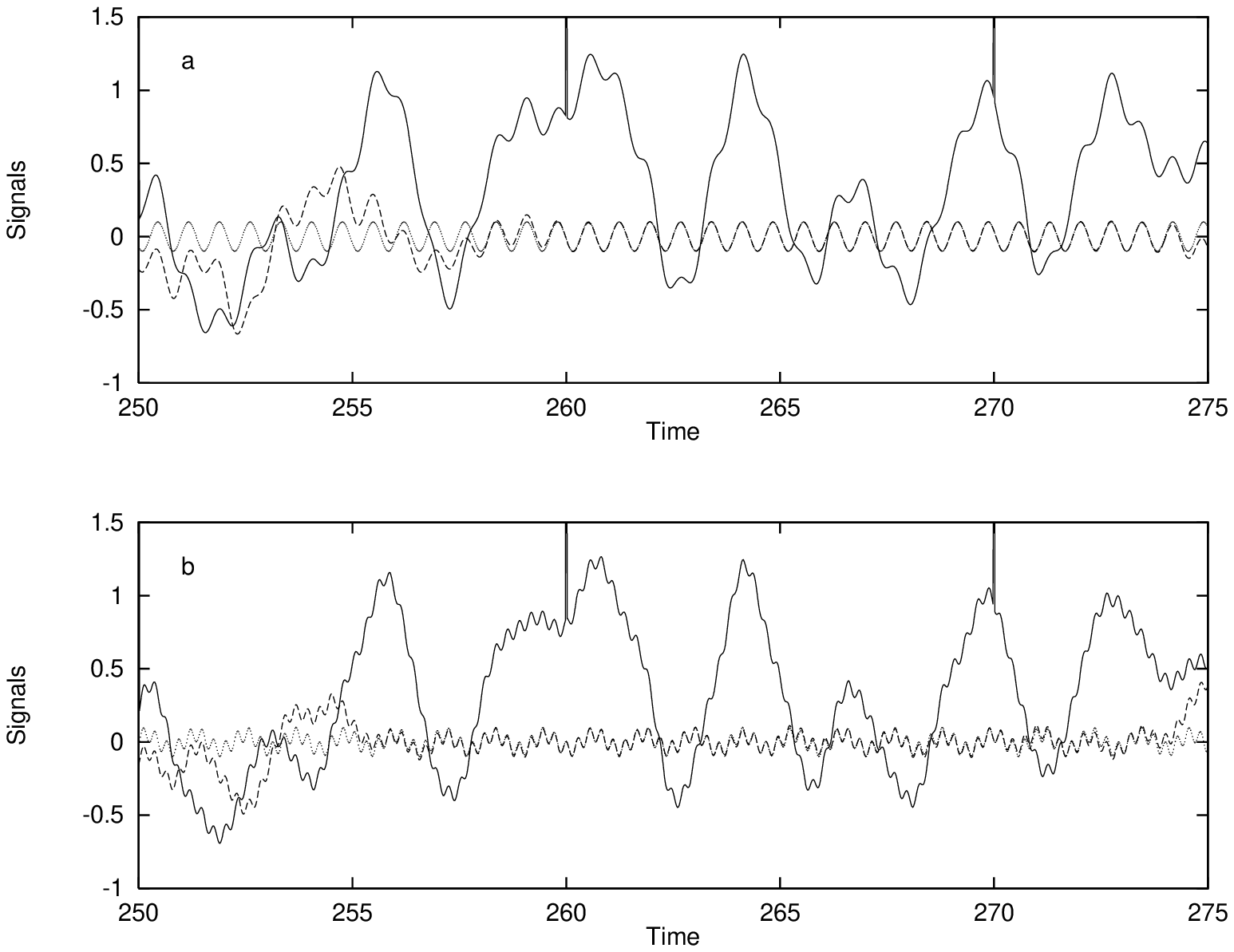}
\end{figure}  
\vskip6cm
Fig.~3 Demonstration of the viability of the secure communication scheme. a)
The signal being communicated is $m(t) = 0.1 \sin \omega_2 t$ (dotted
line) which is added to the output from the SNA, {\it i.e.} $x^{\prime}
(t)$ (solid line). The parameter $q_2$ of the response system differs
from that of the drive by 10\%. Other SNA parameters are as in Fig.~1a
and the recovered signal is the dashed curve.  b) The signal being
communicated is $m(t) = 0.1 \sin \omega_2 t \sin \omega_1 t$ (dotted
line) which is added to the output from the SNA (solid line). The
frequency $\omega_2$ of the response system has fluctuations, with
$\sigma = $10$^{-3}$. Other SNA parameters are as in Fig.~1a. The
recovered signal is the dashed curve. The $\delta$--function spikes in
$x^{\prime} (t)$ are used by the response system for phase matching.

\end{document}